\documentclass[aip,apl,amsmath,reprint]{revtex4-1}
\usepackage{graphicx}

\begin{document}
\title{Influence of tetragonal distortion on the magnetic and electronic properties of the Heusler compound Co$_2$TiSn from first principles}
\author{Markus Meinert}
\email{meinert@physik.uni-bielefeld.de}
\author{Jan-Michael Schmalhorst}
\author{G\"unter Reiss}
\affiliation{Thin Films and Physics of Nanostructures, Department of Physics, Bielefeld University, 33501 Bielefeld, Germany}
\date{\today}

\begin{abstract}
Using the full potential linearized augmented plane wave plus local orbitals method we determine \textit{ab-initio} the lattice parameters of tetragonally distorted Co$_2$TiSn in the L2$_1$ structure. The tetragonal lattice parameter c is determined as a function of the lattice parameter a by energy minimization. The change in total energy is found to be only a few $k_B T$ with respect to room temperature. The spin polarizations as well as the magnetizations are stable against small lattice distortions. It is shown, that the volume is not constant upon distortion and that the volume change is related with significant changes in the magnetization and the gap energy. 

\end{abstract}

\maketitle

Co$_2$ based Heusler compounds have been subject of extensive studies in the context of spintronics during the last decade. They are particularly of interest because of their predicted full spin polarization at the Fermi edge, making them so-called ferromagnetic half-metals. Fully epitaxial magnetic tunnel junctions grown on Cr buffered MgO(001) with a Co$_2$ based Heusler compound as one electrode can exhibit more than 200\,\% tunnel-magnetoresistance ratio (TMR) at room temperature (RT); e.g., 386\,\% with Co$_2$FeAl$_{0.5}$Si$_{0.5}$ \cite{Tezuka09} or 216\,\% with Co$_2$MnSi \cite{Tsunegi09} were demonstrated. Fully epitaxial junctions without Cr buffer exhibiting high TMR of 170\,\% at RT and two Co$_2$MnSi electrodes, \cite{Ishikawa08} as well as junctions with an epitaxial Co$_2$MnGe electrode with 160\,\% TMR at RT are reported. \cite{Taira09}
The \{011\} lattice spacing of MgO differs by a few percent from the typical Heusler \{001\} spacings. Thus, one expects to find significant strain in thin epitaxial films of these compounds on MgO if they are not buffered by a material with intermediate lattice parameter (as Cr). Due to symmetry, a tetragonal distortion is the most likely one. By assuming a constant volume of the unit cell, Block et al. have calculated the spin polarization of Co$_2$CrAl under tetragonal distortions. \cite{Block04} They find that the spin polarization will remain stable if the distortions are of the order of 2-3\,\%.

Here, we present band structure calculations of Co$_2$TiSn (CTS) in the L2$_1$ structure under tetragonal distortions without the assumption of a constant unit cell volume. There are numerous theoretical studies of CTS using diverse band structure schemes. \cite{Ishida82, Mohn95, Galanakis02, Lee05, Hickey06, Miura06, Kandpal07} The results with respect to magnetization and half-metallicity differ considerably with the applied method. Kandpal et al. and Miura et al. predict CTS to be a promising low-moment (2\,$\mu_B$) ferromagnetic half-metal with a high stability against the Co-Ti swap disorder. \cite{Miura06, Kandpal07}

For this study the \textsc{Elk} package, \cite{elk} an all-electron full-potential linearised augmented-plane wave plus local orbitals (FP-LAPW+lo) code, was used. All calculations were performed with a k-point grid of $16\times 16\times 16$ $k$-points on the irreducible wedge of the Brillouin zone, yielding 349 $k$-points. In all calculations the primitive tetragonal cell was used. A smearing width of 0.027\,eV was set and the number of initial empty states was fixed to 40. For the exchange-correlation the generalized gradient approximation (GGA) was employed in the Perdew-Burke-Ernzerhof (PBE) implementation. \cite{PBE96} Spin-orbit coupling of the valence states was neglected, the muffin-tin radii were automatically optimized to be nearly touching. The total energy was converged to $2.7 \cdot 10^{-3}$\,eV. The maximum length of the $\textbf{G} + \textbf{k}$ vectors was given by the muffin-tin radius $R_\text{min}^\text{MT}$ via $R_\text{min}^\text{MT} \times \max{\left|\textbf{G} + \textbf{k}\right|} = 7$. For some critical geometries the total energy and the magnetic moment were checked for convergence with respect to the cut-offs for the basis set; the parameters used for our calculations were found to provide well converged results.

\begin{figure}[b]
\includegraphics[width=8cm]{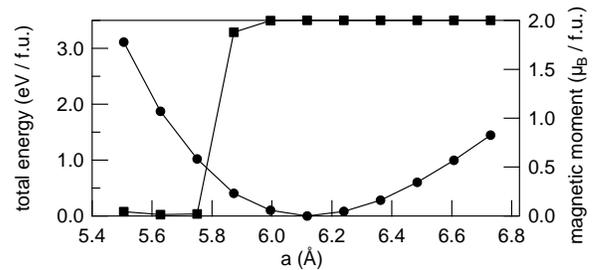}
\caption{Relaxation of the cubic unit cell with total energy (circles) and magnetic moment (squares). The range of lattice parameters corresponds to $\pm$ 10\,\% strain in 2\,\% steps. For the minimum determination more points were calculated, which are not shown here.}
\label{Fig1}
\end{figure}

First, the cubic equilibrium lattice parameter $a_{eq}$ was determined by total energy minimization (Figure \ref{Fig1}). We find 6.12\,\AA{} from theory, whereas the experimentally reported lattice parameter for bulk Co$_2$TiSn is 6.0718\,\AA{}. \cite{Kandpal07} This corresponds to an error of 0.8\,\%. However, it is a common finding, that lattice parameters predicted from density functional theory energy minimizations using the GGA exchange-correlation tend to overestimate the real lattice parameter slightly, whereas local spin density approximation (LSDA) calculations without gradient corrections tend to underestimate them. \cite{Perdew08}

Biaxial strain was simulated by tetragonal lattice distortions using $a = \lambda \cdot a_{eq}$ with $\lambda = 0.94, 0.96,\dots , 1.04, 1.06$. The equilibrium value $c(\lambda)$ was then determined by total energy minimization with fixed $a$. For every $\lambda$ considered, the calculations were performed on 21 values of $c$, making a total of 147 data points.

\begin{figure}[t]
\includegraphics[width=8cm]{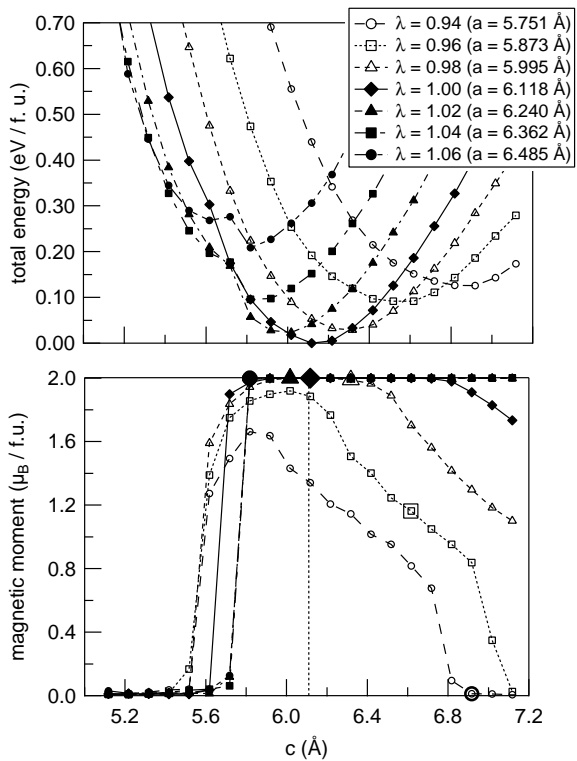}
\caption{Total energy (\textit{top}) and magnetic moment (\textit{bottom}) as a function of the tetragonal lattice parameter c, displayed for the different distortion parameters $\lambda$. The energy is set to zero at the global energy minimum, i.e., at the total energy of the relaxed cubic configuration with $a_\text{eq} = 6.12$\,\AA{}. The magnetic moments corresponding to $E_\text{min}(\lambda,c)$ are tagged with large markers.}
\label{Fig2}
\end{figure}

The total energies and the magnetic moments obtained from the calculations are displayed in Figure \ref{Fig2}. It is immediately clear that the cubic ($\lambda = 1.00$) configuration is the most stable state with minimum energy.

The total energy for biaxial strain of $\pm$\,2\% is increased by 25 - 30 \,meV, i.e., about the typical thermal energy at room temperature. Compared to the energies involved with homogenous lattice compression/expansion (Figure \ref{Fig1}), this energy scale is rather small. Further, the minima of the $\lambda \approx 1$ configurations are relatively broad. This allows for further distortion of the lattice when additional influences get involved. In the case of a thin film growing on a single crystalline substrate, these are, among others, the surface bonding of the substrate and the relaxation of the film at its surface (and, in case of sputtering, interactions with the sputter gas). In fact, the structure has to be considered as very soft with respect to small non-relaxed tetragonal distortions, allowing thin films of CTS to attain very different real structures in dependence on their growth conditions, post-growth treatments and long-range structure. According to our results, a change in the lattice parameters of a CTS thin film with varying growth conditions or annealing temperature cannot solely be ascribed to chemical ordering and recrystallization, but also to strain caused by epitaxial matching with the substrate.

\begin{figure*}[t]
\includegraphics[width=17cm]{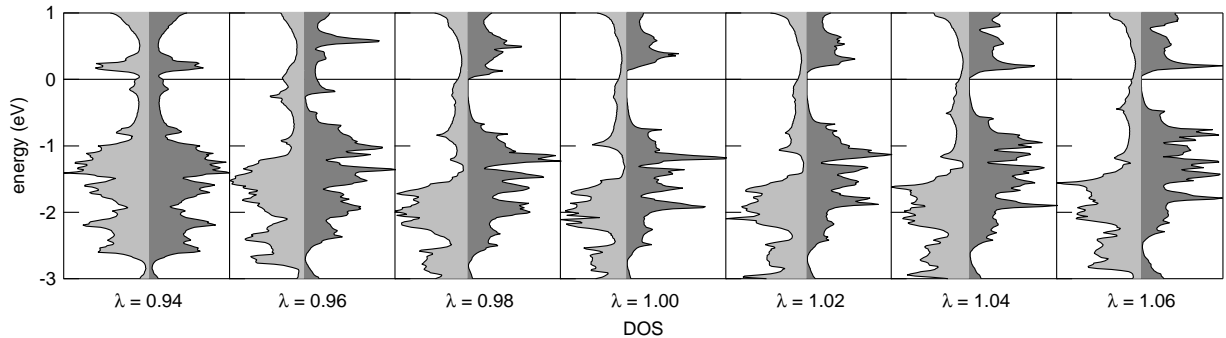}
\caption{Densities of states for all $\lambda$-values with relaxed $c$. Spin-up is always presented on the left, spin-down on the right side of the plots. The Fermi energy is at zero energy.}
\label{Fig3}
\end{figure*}

The magnetic moments in dependence of $c(\lambda)$ reveal a complex response of the density of states and the exchange interaction to the distortions. Here, we take a magnetic moment of 2\,$\mu_B$ / f.u. as indicative for a half-metallic state, which is confirmed by the density of states (Figure \ref{Fig3}). The most striking aspect here is the different behaviour if the lattice is epitaxially compressed ($\lambda < 1$) or expanded ($\lambda > 1$). With increasing compression the half-metallicity vanishes quickly or even magnetism is lost completely ($\lambda = 0.94$). In the case of epitaxial expansion the situation is different, half-metallicity is conserved here. The discontinuous transition from the ferromagnetic to the non-magnetic case with decreasing $c$ pins the system in the ferromagnetic, half-metallic state.
The reason for the loss of the full magnetic moment for the most non-relaxed geometries considered here is a closing of the gap, which is due to the breaking of the cubic crystal field symmetry. This in turn is responsible for the formation of common bonding and anti-bonding hybrid bands, which drive the gap formation. \cite{Galanakis02} The closing of the gap can not be ascribed to any atomic site in particular; Co, Ti as well as Sn exhibit a non-zero DOS at the position of the gap when the magnetization is less than 2\,$\mu_B$/f.u.

A sharp transition from magnetic to non-magnetic is observed if $c$ is reduced below a critical value $c_\text{crit}$, which depends weakly on $\lambda$: in the compressed case, $c_\text{crit}$ is a little smaller than for $\lambda = 1$, whereas it is a little bit larger in the expanded case. This is clear, considering that the geometry is closer to the cubic for a smaller $c$ with $\lambda < 1$ and for a larger $c$ with $\lambda > 1$. It is remarkable that $c_\text{crit}$ does not depend on the actual value of $\lambda$ (on the scale of our $c(\lambda)$ sampling). However, this critical lattice parameter of 5.5 - 5.8\,\AA{} is also found for the homogenous compression of the lattice (Figure \ref{Fig1}), indicating a nearly geometry-independent minimum distance for the ferromagnetic coupling of the Co atoms.

\begin{figure}[b]
\includegraphics[width=8cm]{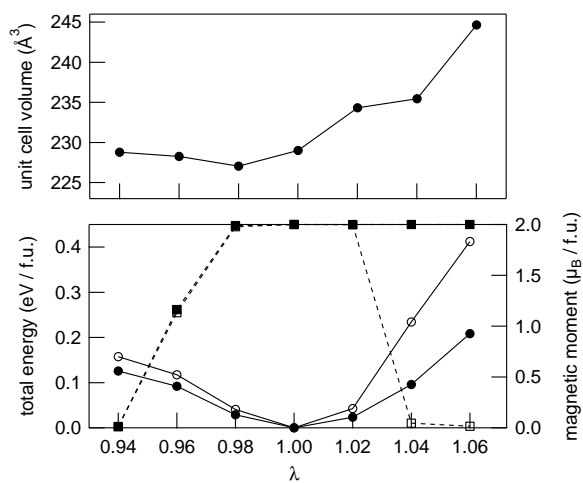}
\caption{\textit{Top:} Relaxed unit cell volume as a function of the distortion parameter $\lambda$. \textit{Bottom:} Total energy (circles) and magnetic moment (squares) for constant volume (open symbols) and relaxed (filled symbols) calculations.}
\label{Fig4}
\end{figure}

The comparison of distortion calculations with constant volume and the relaxed calculations (Figure \ref{Fig4}) reveals that the assumption of a constant volume is only partially justified. For $\lambda < 1$ it turns out, that the volume remains essentially constant. In contrast, for $\lambda > 1$ the volume increases significantly. For $\lambda = 1.06$ we observe a volume increase of 6.8\,\%. This large discrepancy is caused by the above mentioned energetic pinning to the ferromagnetic state. Here we also see the largest difference between the constant-volume calculations and the fully relaxed ones: whereas the constant-volume calculations predict a very small window of $\lambda$ in which the compound is magnetic, the fully relaxed calculations rather predict a quite large stability of the ferromagnetism and even the half-metallic gap.

In summary we have shown by \textit{ab-initio} band structure calculations that the cubic crystal structure of Co$_2$TiSn allows easily for tetragonal distortion, which we therefore expect to occur in epitaxially grown thin films. The half-metallicity is retained over a large range of distortion. We provide an additional explanation for the typically observed large spectrum of lattice parameters of Heusler compound thin films in dependence on deposition and post-treatment conditions. In comparison with calculations that assume a constant volume, we show that this assumption is not justified due to the energetically favorable ferromagnetism. Thus, when dealing with distortions in Heusler compounds, it is necessary to consider relaxed geometries. We assume, that these findings are representative for other Co$_2$ based Heusler compounds, although the results may vary in detail. Further investigation on this subject is necessary. 

This work was supported by the German Bundesministerium f\"ur Bildung und Forschung (BMBF). We thank Dr. John Kay Dewhurst for helpful advice on running the \textsc{Elk} package.

\newpage
\end{document}